\documentstyle[12pt]{article}
\begin{document}
\begin{center}
{\bf On  Quantum Communication Channels with Constrained
Inputs}
\vskip20pt
A. S. Holevo\footnote{Visiting Research Scholar at the Department of
Electrical and Computer Engineering, Northwestern University.
}\\Steklov Mathematical Institute, Moscow. 
\end{center}\vskip30pt
{\small {\bf Abstract} -- The purpose of this work is to extend 
results of previous papers \cite{hol}, \cite{bur} 
to quantum channels with additive
constraints onto the input signal, by showing that the capacity of such 
channel is equal to the supremum of the
entropy bound with respect to all apriori distributions satisfying the
constraint.  We also make an extension to channels with continuous
alphabet. As an application, we prove the formula for the capacity of 
the quantum Gaussian channel
with constrained energy of the signal, establishing the asymptotic
equivalence of this channel to the quasiclassical photon channel, and
derive the lower bounds for the reliability function of the
pure-state Gaussian channel.}
\vskip30pt
\centerline{\sc I. The case of discrete alphabet}
\vskip10pt

Most of the results concerning the capacity of quantum
communication channels were proved for the case of finite input
alphabets \cite{hol73}, \cite{hol79}, \cite{jozsa}, \cite{hol}. The
importance of considering channels with infinite (continuous)
alphabets and with constrained inputs was clear
from the beginning of quantum communications \cite{gordon},
\cite{le}, and was
reiterated in \cite{yuen}. 
The present paper, which is a continuation of our papers
\cite{hol}, \cite{bur}, is devoted to study of this case.

For reader's convenience we start with repeating some basic notions
from \cite{hol}, making necessary modifications for channels with
constrained inputs.  Let $\cal H$ be a
 Hilbert space.   A quantum
communication channel with (possibly infinite) discrete input 
alphabet $A = \{i\}$ consists of the mapping
$i\rightarrow S_i$ from the input alphabet to the set of density
operators (d. o.) in
$\cal H$.  The input is described by an {\sl apriori}
probability distribution $\pi = \{\pi_i\}$ on $A$. At the output there is a
quantum measurement in the sense of \cite{hol73}, given by
{\sl resolution of identity} in $\cal H$, that is by
 a family $X = \{ X_j\}$ of positive
operators in $\cal H$ satisfying $\sum_j X_j = I$, where $I$ is the unit
operator in $\cal H$. The index $j$ runs through some discrete output alphabet,
which is not fixed here.  The conditional probability of the output $j$ if the
input was $i$ equals to $P(j|i) = \mbox{Tr}S_i X_j$. The Shannon information is
given by the classical formula 
\begin{equation} \label{1}
I_1 (\pi , X ) = \sum_j \sum_i \pi_i
P(j|i)\mbox{log}\left( \frac{P(j|i)} {\sum_k \pi_k P(j|k)}\right) .
\end{equation}

Let $f(i)$ be a function defined on the input alphabet. We shall
restrict the apriori distributions $\pi$ by the inequality
\begin{equation} \label{2}
\sum_i f(i) \pi(i) \leq E ,
\end{equation}
where $E$ is a real number, and denote the class of such probability
distributions by ${\cal P}_1$. 

Let us consider also the product channel in the 
Hilbert space ${\cal H}^{\otimes n} =
{\cal H}\otimes ...\otimes {\cal H}$ with the input alphabet $A^n$ consisting
of words $u = (i_1 ,...,i_n )$ of length $n$, and the d. o.  $$S_u =
S_{i_1}\otimes ...\otimes S_{i_n} $$ corresponding to the word $u$.
If ${\pi}$ is a probability distribution on $A^n$ and $X$ is a resolution of
identity in ${\cal H}^{\otimes n}$, we define the information quantity $I_n
(\pi , X )$ by the formula similar to (\ref{1}).  
We put the additive constraint onto the distribution $\pi$ by asking
\begin{equation} \label{3}
\sum_{i_1, \ldots, i_n} [f(i_1)+\ldots + f(i_n)] \pi (i_1,\ldots,i_n)
\leq nE ,\end{equation}
and denote by ${\cal P}_n$ the class of probability distributions
satisfying this constraint.

Defining 
\begin{equation} \label{sup}
C_n = \sup_{\pi\in{\cal P}_n;\, X} I_n
( \pi ,X),\end{equation}
we have the property of superadditivity $C_n + C_m \leq C_{n+m}$,
hence the following limit exists 
\begin{equation} \label{4}
C = \lim_{n \to \infty}C_n /n , \end{equation}
which we call the {\sl capacity} of the initial channel. (This
quantity may be eventually infinite for 
infinite dimensional Hilbert space and in what follows we consider the
nontrivial case $C < \infty .$). The
definition is justified by the classical Shannon's
coding theorem for channels with constrained inputs.
Namely,  call by {\sl code of size} $N$ a sequence $(u_1 , X_1 ),..., (u_N
, X_N )$, where $u_k$ are words of length $n$, and $\{ X_k \}$
is a family of
positive operators in ${\cal H}^{\otimes n}$, satisfying $\sum_{j=1}^N X_j \leq
I$. Defining $X_0 = I - \sum_j X_j$, we have a resolution of identity in
${\cal H}^{\otimes n}$.  An output $k \geq 1$ means decision that the
word $u_k$ was transmitted, while the output $0$ is interpreted as evasion of
any decision.The average error probability for such a code is 
\begin{equation} \label{4a}{\bar \lambda} =
\frac{1}{N} \sum_{k=1}^N [1 - \mbox{Tr}S_{u_k} X_k ]. \end{equation} 
Let us denote $p(n,
N)$ the minimum of this error probability with respect to all codes of the size
$N$ with words of length $n$, satisfying the condition
\begin{equation} \label{5}
f(i_1)+ \ldots + f(i_n)\leq nE .\end{equation}
As a direct consequence of theorems in Sec. 7.3 of \cite{gal}, one can
prove the following statement providing information-theoretic
justification of the definition (\ref{4}):

{\bf Proposition 1.} {\sl If $R < C$ , then $p(n, 
\mbox{e}^{nR}) \rightarrow 0$ ,
and if} $R > C$ ,~~ {\sl then}~~~
$\\ p(n, \mbox{e}^{nR})\not\rightarrow 0$.

Let
$H(S) = - \mbox{Tr}S\mbox{log}S$ be the von Neumann entropy of a d. o.  $S$
and let $\pi = \{\pi_i \}$ be an apriori distribution on $A$. We  denote
$${\bar S}_{\pi} = \sum_{i\in A}\pi_i S_i $$ and assume that 
\begin{equation} \label{a5}
\sup_{\pi\in{\cal P}_1} H( {\bar S}_{\pi}) < \infty .\end{equation}
Further, we denote \begin{equation} \label{5a}
\Delta H (\pi ) = H({\bar S}_{\pi}) - 
\sum_{i\in A}\pi_i H(S_i ) .\end{equation} 

{\bf Proposition 2.} {\sl 
Under the condition} (\ref{a5}) \begin{equation} \label{6}
C = \sup_{\pi\in{\cal P}_1} \Delta H(\pi ) .\end{equation}

{\sl Proof}. To prove the~ $\leq$~ part, take $\pi \in {\cal P}_n$. 
The {\sl entropy bound} \cite{hol73}, \cite{yuen} implies
$$I_n (\pi, X) \leq \Delta H_n (\pi ) ,$$ 
where $\Delta H_n (\pi )$ is the analog of $\Delta H (\pi )$ for the
product channel.
According to the subadditivity
property of quantum entropy \cite{araki}, $$\Delta H_n (\pi )\leq
\sum_{k=1}^n \Delta H (\pi^{(k)} ),$$ where $\pi^{(k)}$ is the $k$-th
marginal distribution of $\pi$ on $A$. Therefore $$ {1 \over n}
I_n (\pi, X) \leq {1 \over n}\sum_{k=1}^n \Delta H (\pi^{(k)} )
\leq \Delta H ({\bar \pi}),$$ where ${\bar \pi} = {1 \over
n}\sum_{k=1}^n \pi^{(k)}$, since $\Delta H (\pi )$ is concave
function of $\pi$ \cite{hol73}. Also inequality (\ref{3}) can be rewritten as
$$\sum_{k=1}^n \sum_{i_k}f(i_k ) \pi^{(k)}(i_k) \leq nE .$$
It follows that $n^{-1} C_n \leq \sup_{\pi\in{\cal P}_1} \Delta H(\pi )$ and
hence, a similar inequality holds for $C$ .

To prove the~ $\geq$~ part, we use the  random coding modified for constrained
inputs. Let $\pi$ be a distribution satisfying (\ref{2}), and let {\sf
P} be a distribution on the set of $M$ words, under which the words
are independent and
$${\sf P} (u = (i_1, \ldots, i_n)) =
\pi_{i_1}\cdot\ldots\cdot\pi_{i_n}.$$
Let $\nu_n = {\sf P} ({1 \over n}\sum_{k=1}^n f(i_k) \leq E)$ and define the
modified distribution 
$${\tilde {\sf P}} (u = (i_1,\ldots,i_n)) = 
\left\{\begin{array}{ll}
\nu_n^{-1}  \pi_{i_1}\cdot\ldots\cdot\pi_{i_n}, & \mbox{if}\,
\sum_{k=1}^n f(i_k)\leq nE,\\0, &
\mbox{otherwise.}\end{array} 
\right.$$ Let us remark that if $\pi\in {\cal P}_1$, then ${\sf M}f
\leq E$ (where {\sf M} (${\tilde {\sf
M}}$) is the expectation corresponding to {\sf P} (${\tilde {\sf
P}}$)) and hence by the central limit theorem
$$\lim_{n\rightarrow\infty}\nu_n\geq
1/2 .$$ Therefore 
${\tilde {\sf M}} \xi \leq 2^m {\sf M} \xi$ for any nonnegative random
variable $\xi$ depending on $m$ words.  In particular, for the error
probability (\ref{4a}) we gave in \cite{hol} the upper bound (17)
depending on arbitrary two words, the expectation of which with
respect to  {\sf P} can be made arbitrarily small provided  
$M=\mbox{e}^{nR}, n\rightarrow\infty,$ with
$R < C$. Thus ${\tilde {\sf M}}{\bar \lambda}$ also can be made
arbitrarily small under the same circumstances.
The proof in \cite{hol} is for finite dimensional Hilbert
space, but under the condition (\ref{a5}) it can be modified for
infinite dimensions. 
Since the distribution ${\tilde{\sf P}}$ is concentrated on
words satisfying (\ref{5}), we can choose a code satisfying this
constraint for which ${\bar \lambda}$ can be made arbitrarily small.
Proposition 1 then implies that $$C \geq \sup_{\pi\in{\cal P}_1} 
\Delta H(\pi ), $$ which completes the proof.
\vskip20pt 
\centerline{\sc II. The case of continuous alphabet} 
\vskip10pt

In this section we take as the input alphabet $A$ arbitrary Borel subset
in a finite-dimensional Euclidean space $\cal E$.  We assume that the
channel is given by {\sl weakly continuous}
 mapping $x\rightarrow S_x$
from the input alphabet $A$ to the set of
density operators in $\cal H$. (The weak continuity means continuity
of all matrix elements $<\psi |\, S_x\, \phi_ >;  \psi,\phi \in \cal H$). 
We assume that
a {\sl continuous} function $f$ on $\cal E$ is fixed and consider the
set ${\cal P}_1$ of probability measures $\pi$ on $A$ satisfying
\begin{equation} \label{7}
\int_{A} f(x) \pi (dx) \leq E .
\end{equation}

 Like in the classical case, we
discretize the channel by taking apriori distributions with discrete
supports 
\begin{equation} \label{8}
\pi(dx) = \sum_{i}\pi_i \delta_{x_i} (dx),\end{equation} where $\{
x_i\}\subset A$ is arbitrary countable collection of points and $$
\delta_x (B) =\left\{\begin{array}{ll}1, & \mbox{if}\quad x\in B,\\
0, & \mbox{if} \quad x \not\in B,\end{array}\right.$$
and by taking discrete resolutions of identity
$\{ X_j \}$. For $\pi$ of the form (\ref{8}) the constraint (\ref{7})
takes the form (\ref{2}). Then we define the capacity of the channel
$x\rightarrow S_x$ with the constraint (\ref{7}) by repeating 
the argument in Sec. 1
with the only modifications that
$${\sf P}(j|i) = \mbox{Tr} S_{x_i} X_j ,$$ and
additional supremum in (\ref{sup}) 
is taken over all possible choices of the points $x_i \in A$.

For arbitrary $\pi \in {\cal P}_1$ consider the quantity
\begin{equation} \label{9}
\Delta H (\pi ) = H ({\bar S}_{\pi}) - \int_A H (S_x ) \pi (dx) ,
\end{equation}
where\begin{equation} \label{9a}
{\bar S}_{\pi} = \int_A S_x \pi (dx).
\end{equation}
Because of the weak continuity of the function $S_x$ the integral
is well defined and represents a density
operator in $\cal H$. Moreover, from the Lemma below it follows 
that the nonnegative 
function $H(S_x )$ is lower semicontinuous, and hence the second term in
(\ref{9}) is also well defined . 

{\bf Proposition 3}. {\sl Under the condition (\ref{a5}), in which
${\bar S}_{\pi}$ is given by (\ref{9a}) and ${\cal P}_1$ -- by
(\ref{7}),
\begin{equation}
\label{11} C = \sup_{\pi \in {\cal P}_1} \Delta H (\pi ) . \end{equation}

Proof. } The~~ $\leq$~~ part of the proof follows obviously from
Proposition 2, as for $\pi$ given by (\ref{8}) the quantity (\ref{9})
turns into (\ref{5a}). To prove the~~ $\geq$~~ part it is sufficient to
construct, for arbitrary $\pi\in {\cal P}_1$, a sequence of discrete
${\pi}^{(l)} \in {\cal P}_1$ such that
\begin{equation} \label{12}
\lim\inf_{l\rightarrow \infty} \Delta H (\pi^{(l)}) \geq \Delta H (\pi
) .\end{equation}
To this end for any $l =1, \ldots$ we consider the division of
$A$ into disjoint subsets
\begin{equation} \label{13}
B_k^{(l)} = \{ x: k/l\leq H(S_x )<(k+1)/l \},\quad k=\ldots, -1, 0,
1, \ldots\, .\end{equation} By making, if necessary, a 
finer subdivision, we can
always assume that diameters of all sets $B_k^{(l)}$ are bounded from
above by $\epsilon_l$, where $\epsilon_l\rightarrow 0$ as
$l\rightarrow\infty$.
Let $x_k^{(l)}$ be a point at which $f(x)$ achieves its
minimum on the closure ${\bar B_k^{(l)}}$ of $B_k^{(l)}$, and define
\begin{equation} \label{13a}
\pi^{(l)} (dx) = \sum_k \pi (B_k^{(l)}) \delta_{x_k^{(l)}} (dx) ,
\end{equation}
where $\pi$ is a fixed distribution from ${\cal P}_1$.
Then $$\int_A f(x) \pi^{(l)}(dx) \leq \int_A f(x) \pi (dx),$$ hence
$\pi^{(l)}\in {\cal P}_1$.

By construction (\ref{13}), (\ref{13a}) 
and due to the condition (\ref{a5}) we have
$$\left|\int_A H(S_x ) \pi^{(l)} (dx) - \int_A H(S_x ) \pi (dx)\right| \leq
1/l ,$$ and it remains to show that \begin{equation} \label{14}
\lim\inf_{l\to\infty} H\left(\int_A S_x \pi^{(l)} (dx)\right) \geq
H\left(\int_A S_x \pi (dx)\right) .\end{equation}
 
We first remark that due to the weak continuity and uniform
boundedness of the function $S_x$, the density operators $\int_A S_x
\pi^{(l)} (dx)$ weakly converge to the density operator $\int_A S_x
\pi (dx)$. Indeed, let $B_c$ be the ball of radius $c$ in ${\cal E}$.
Then $$\left| <\phi | \int_A S_x
\pi^{(l)} (dx) \psi> - <\phi | \int_A S_x
\pi (dx) \psi> \right| $$ $$\leq \sum_k \int_{B_k^{(l)}\cap B_c}
|<\phi | S_{x_k^{(l)}}\psi> - <\phi |S_x \psi>|
\pi (dx) + 2\|\phi\| \|\psi\| \pi (A\setminus B_c) .$$ By choosing first $c$
large enough to make the second term small, we can make the first term
small for all large enough $l$ since $<\phi | S_x \psi>$ is uniformly 
continuous on $A\cap B_c$ and the diameters of $B_k^{(l)}$ uniformly
tend to zero.

It remains to apply the following Lemma (this result is well known but
we include sketch of its proof for completeness): 

{\bf Lemma}. {\sl Let $\{ A_l \}$ be a sequence of density operators,
weakly converging to a density operator $A$. Then
$$\lim\inf_{l\rightarrow\infty} H(A_l ) \geq H(A) .$$

Proof.} Let $\{ P_m \}$ be a monotonely increasing sequence of
finite-dimensional projections weakly converging to unit operator in
$\cal H$. By Lemma 4 from \cite{lin} the sequence $$H(P_m S P_m) +
\mbox{Tr}P_m S P_m \log \mbox{Tr}P_m S P_m$$ monotonely converges to $
H(S)$ for any d. o. $S$. Then  we have
$$\lim\inf_{l\to\infty} H (A_l ) = \lim\inf_{l\to\infty}\,    
\lim_{m\to\infty} H( P_m A_l P_m ) \geq \lim_{m\to\infty}\lim_{l\to\infty}
H(P_m A_l P_m ) = H(A) .$$

This completes the proof of (\ref{12}) and hence of Proposition 3. 
\vskip20pt
\begin{center}{\sc III. The quantum Gaussian channel\\ with constrained
energy of the signal}\end{center}

 Let $A$ be the complex plane {\bf C}, and let for every
 $\alpha  \in
{\bf C}$ the density operator $S_\alpha $ describe the thermal state of
harmonic oscillator with the signal amplitude $\alpha $ and the mean number
of the noise quanta $N$, i. e.  
\begin{equation} \label{15}
S_\alpha  ={1 \over \pi N} \int\mbox{exp}\left( - {|z-\alpha |^2 \over
N}\right) |z><z| d^2 z ,\end{equation} where $|z>$ are the coherent
state vectors. This is quantum analog of
channel with additive Gaussian noise (see \cite{gordon}, \cite{hel},
\cite{hol77}, \cite{caves}). 
We remind for future use that $$S_{\alpha} = V(\alpha ) S_0 V(\alpha
)^*,$$ where $$V(\alpha ) = \mbox{exp}(\alpha a^{\dagger} - {\bar
\alpha}a)$$ are the unitary displacement operators, $a^{\dagger},
a$ being the creation - annihilation operators for the harmonic
oscillator, and the operator $S_0$ has 
the spectral representation:
\begin{equation}\label{29}
S_0 ={1 \over N+1}  \sum_{n=0}^{\infty}\left({N \over N+1}\right)^n |n><n|,
\end{equation}
where $|n>$ are the eigenvectors of the number operator $a^{\dagger}a$.
The states (\ref{15}) all have the same entropy
\begin{equation} \label{21}
H ( S_\alpha  ) = (N+1)\log (N+1) - N\log N ,\end{equation}
and the mean
number of quanta 
\begin{equation} \label{22}
\mbox{tr} S_{\alpha}\, a^{\dagger}a = N + |\alpha |^2 .
\end{equation}

We impose the following
constraint onto the mean energy of the signal
\begin{equation} \label{15a}\int |\alpha |^2 \,\pi (d^2 \alpha ) \leq
E ,\end{equation} where $\pi (d^2 \alpha )$ is an apriori distribution.
In fact, $E$ is the ``mean number of quanta'' in the signal, which is
proportional to energy for one mode.
Consider the density operator $${\bar S}_{\pi} = 
\int S_\alpha\,  \pi (d^2 \alpha ) .$$ The
constraint (\ref{15a}) by virtue of (\ref{22}) 
implies \begin{equation} \label{23}
\mbox{Tr} \, {\bar S}_{\pi}\, a^{\dagger} a \leq N+E .
\end{equation}
It is well known that under this constraint the maximal entropy 
\begin{equation} \label{24a}
H( {\bar S}_{\pi}) = (N+E+1)\log (N+E+1) - (N+E)\log (N+E)
\end{equation}
is attained by Gaussian density operator
\begin{equation} \label{25}
{\bar S}_\pi  ={1 \over \pi (N+E)} \int\mbox{exp}\left( - {|z|^2 \over
(N+E)}\right) |z><z| d^2 z ,\end{equation}
 corresponding to the optimal apriori distribution 
\begin{equation} \label{opt}\pi
(d^2 \alpha ) = {1 \over \pi E}\, \mbox{exp}\left( - {|\alpha |^2 \over
E}\right) d^2 \alpha .\end{equation}
Hence the condition of Proposition 3 is fulfilled, and 
the capacity of the channel is equal to
$$C = H( {\bar S}_{\pi}) - H ( S_\alpha  ) = \log \left(1 + {E \over
N+1}\right)$$ $$ + (N+E)\log \left(1 + {1 \over
N+E}\right) - N \log \left(1 + {1 \over N}\right).$$

This quantity was anticipated in \cite{gordon} (relation (4.28) )
as an upper bound for the information transmitted by the quantum
Gaussian channel. On the other hand, for a long time this quantity was
also known as the capacity of the  ``narrow band photon
channel'' \cite{gordon62}, \cite{le}. Our argument based on
Proposition 3 gives for the first time 
the proof of the asymptotic equivalence, in the
sense of information capacity, of the
Gaussian channel with the energy constraint (\ref{15a}) to this
quasiclassical channel. To make the point clear, we give below a simplified
one-mode description of the  photon channel.

Consider the discrete family of states
\begin{equation} \label{30}
S_m = P( m ) S_0 P( m )^*,\qquad m = 0, 1,\dots , \end{equation}
where $P(m)$ is energy shift operator satisfying $ P(m)|n>
=|n+m>.$ Notice that $P( m ) = P^m$, where $P$ is 
isometric operator adjoint to the quantum-mechanical 
``phase operator'' \cite{prob}. The states $S_m$ all have
the same entropy (\ref{21}) as the states $S_{\alpha}$, and the mean
number of quanta 
\begin{equation} \label{31}
\mbox{tr} S_m \, a^{\dagger}a = N + m .
\end{equation}
Moreover, all states (\ref{30}) are diagonal in the number
representation, in which sense the channel may be called
quasiclassical.

Imposing the constraint
\begin{equation} \label{32}
\sum_{m=0}^{\infty} m \, \pi_m \leq E ,\end{equation}
where $\pi_m$ is an apriori distribution, and introducing the density
operator $${\bar S}_{\pi}' = \sum_{m=0}^{\infty} \pi_m S_m, $$
by virtue of (\ref{31}), we obtain the same constraint (\ref{23})
for the new operator ${\bar S}_{\pi}'$. The maximal entropy (\ref{24a})
is again attained by the operator (\ref{25}), which has the spectral 
representation 
\begin{equation}\label{33}
{\bar S}_{\pi} = {1 \over N+E+1}  \sum_{n=0}^{\infty}\left({N+E \over
N+E+1}\right)^n |n><n|.
\end{equation}
It corresponds to the
optimal apriori distribution \cite{le}
$$\pi_m = {N \over N+E} \delta_{m0} + {E \over N+E}\left[{1 \over
N+E+1} \left({N+E \over N+E+1}\right)^m\right] .$$

There is notable difference between the case of pure-state channel as
opposed to
the general case. For a pure-state case (where $N = 0$), one can formulate
a broader problem of finding a maximum capacity channel $x\rightarrow
S_x$ with arbitrary alphabet $\{x\}$ 
and an apriori distribution $ \pi (dx)$ satisfying the output constraint 
$$\mbox{Tr}{\bar S}_{\pi}\, a^{\dagger} a \leq E .$$
This was done in \cite{yuen} where it was shown that the noiseless
photon channel provides a solution to this problem. In view of the
result of \cite{jozsa}, any other pure-state channel satisfying
$$\int S_x \pi (dx) = {1 \over E+1} \sum_{n=0}^{\infty}\left({E \over
E+1}\right)^n |n><n| $$ gives, asymptotically, a solution to the same 
problem.However, in the
general case imposing the output constraint (\ref{23}) instead of the
input constraints (\ref{15a}) or (\ref{32}) looks rather artificial;
the equivalence of these constraints for apparently different channels
seems to be a very special feature of the quantum Gaussian density
operators.

\vskip30pt
\centerline{\sc IV. The upper bounds for error probability}
\vskip10pt

A much more detailed information concerning the rate of convergence of
the error probability can be obtained for pure-state channels, by
modifying the estimates from \cite{bur} to channels with infinite
alphabets and constrained inputs following the method of \cite{gal},
Ch. 7.  We start with the case of discrete alphabet.

Let $S_i = |\psi_i><\psi_i|$ be the pure signal states of the channel,
and let $\pi$ be an apriori distribution satisfying the restriction
(\ref{2}). Then the
following {\sl random coding bound} holds for the error probability $p(n,
N)$ where $N = \mbox{e}^{nR}$ with $R < C$:
\begin{equation} \label{40}
p(n, \mbox{e}^{nR}) \leq 2 \left({\mbox{e}^{p\delta} \over \nu_{n,
\delta }}\right)^2 \mbox{exp}\{-n[{\mu (\pi, s, p) - s R]}\},
\end{equation}
where \begin{equation} \label{41}
\mu (\pi, s, p) = - \log \mbox{Tr}\left\{\sum_i \pi_i \mbox{e}^{p[f(i)
- E]} S_i\right\}^{1+s}, \end{equation}
and $0\leq s\leq 1, 0\leq p, 0 < \delta $ are arbitrary parameters. 
The quantity $$\nu_{n, \delta } =  {\sf P}(En - \delta \leq 
\sum_{k=1}^n f(i_k)\leq nE )$$ satisfies
$\lim_{n\to \infty}\sqrt{n}\nu_{n,\delta } > 0$, thus adding only
$o(n)$ to the exponential in (\ref{40}).

The bound (\ref{40}) is obtained in the same way as Proposition 1 in
\cite{bur}, that is by evaluating the expectation of the average error
probability (\ref{4a}) using random, independently chosen codewords, but
with the modified codeword distribution
\begin{equation} \label{42}
{\tilde {\sf P}}_{\delta} (u = (i_1,\ldots,i_n)) = 
\left\{\begin{array}{ll}\nu_{n,\delta }^{-1}\,  
\pi_{i_1}\cdot\ldots\cdot\pi_{i_n}, & \mbox{if}\, nE - \delta \leq 
\sum_{k=1}^n f(i_k)\leq nE,\\
0, & \mbox{otherwise.}\end{array} \right.\end{equation}
The point is that for any random variable $\xi$ depending on $m$ words
$${\tilde {\sf M}}_{\delta}\,\xi \leq \left({\mbox{e}^{p\delta} \over
\nu_{n,\delta } }\right)^{m}\, {\sf M}\mbox{exp}\{mp\sum_{k=1}^n [f(i_k) -
E]\} \,\xi,$$ where $p\geq 0$.
By using this inequality  after equation (14) in the proof of
Proposition 1 from
\cite{bur}, and following argument in Ch. 7 of \cite{gal}, 
we can obtain the bound (\ref{40}).

In the same way, the proof of Proposition 2 from \cite{bur} can be
modified to obtain the {\sl expurgated bound}
\begin{equation} \label{43}
p(n, \mbox{e}^{nR}) \leq  \mbox{exp}\{-n[{\tilde \mu} (\pi, s, p) - s (R + {2
\over n}\log{2\mbox{e}^{p\delta} \over \nu_{n,\delta}})]\},
\end{equation}
where \begin{equation} \label{44}
{\tilde \mu} (\pi, s, p) = - s \log \sum_{i,k}\pi_i\pi_k\mbox{e}^{p[f(i)
+ f(k) - 2E]} \,|<\psi_i |\psi_k>|^{2/s}.\end{equation}

These bounds can be extended to pure-state channels with continuous
alphabets by using technique of Sec. II to obtain (\ref{40}), (\ref{43}) with
\begin{equation} \label{45}
\mu (\pi, s, p) = - \log \mbox{Tr}\left\{\int_A \mbox{e}^{p[f(x)
- E]} S_x \pi (dx) \right\}^{1+s}, \end{equation}
\begin{equation} \label{46}
{\tilde \mu} (\pi, s, p) = - s \log \int_A\int_A \mbox{e}^{p[f(x)
+ f(y) - 2E]} |<\psi_x |\psi_y>|^{2/s}\pi (dx) \pi (dy).\end{equation}

Introducing the {\sl reliability function}$$
E(R) = \lim\sup_{n\to\infty}{1 \over n}\log {1 \over p(n, \mbox{e}^{nR})},$$
which characterizes the exponential rate of convergence of the error
probability, we get the lower bound for $E(R)$:
$$E(R)\geq \max \{ E_r (R), E_{ex} (R)\},$$
where
\begin{equation} \label{47}
E_r(R) = \max_{ 0\leq s \leq 1} 
(\max_{0\leq p}\max_{\pi\in{\cal P}_1}\mu (\pi, s, p) -sR), \end{equation}
\begin{equation} \label{48}
E_{ex}(R) = \max_{ 1\leq s} 
(\max_{0\leq p}\max_{\pi\in{\cal P}_1}{\tilde \mu} (\pi, s, p) -sR). 
\end{equation} An example where the
maximization at least partially can be performed analytically is
considered in the following Section.
\vskip20pt
\begin{center}{\sc V. The reliability function\\ of quantum Gaussian
pure-state channel}
\end{center}

We are going to apply results of the previous Section to the
Gaussian pure-state channel $\alpha \rightarrow  S_{\alpha} = 
|\alpha><\alpha|$ with the constraint  (\ref{15a}). By  taking the optimal
apriori distribution (\ref{opt}) we can calculate explicitly the
functions (\ref{45}), (\ref{46}).  

Namely, to calculate (\ref{45}), we remark that$$
\int\mbox{e}^{p(|z|^2 - E)} S_z \pi(d^2 z) = {\mbox{e}^{-pE} \over 1-pE}\,
{1 \over \pi E'}\int \mbox{e}^{-{|z|^2 \over E'}} |z><z| d^2 z$$ $$ =
 {\mbox{e}^{-pE} \over 1-pE}\,
{1 \over E'+1}\sum_{n=0}^{\infty}\left({E' \over E'+1}\right)^n |n><n|,$$
where $E' = E/(1-pE)$, provided $p < E^{-1}$, 
and the trace of the $(1+s)$-th power of this operator is easily
calculated to yield
\begin{equation} \label{50}
\mu (\pi, s, p) = (1+s)pE + \log [(1+E - pE)^{1+s} -
E^{1+s}].\end{equation}
 
By taking into account that $$|<z|w>|^2 = \mbox{e}^{-|z-w|^2},$$
(see, e. g. \cite{hel}), we can calculate the integral in (\ref{46}) as
$${\mbox{e}^{-2pE} \over (\pi E)^2}\int\int \exp\{ - [(E^{-1}+s^{-1}-p)|z|^2 
+ (E^{-1}+s^{-1}-p)|w|^2 -2s^{-1}\mbox{Re}{\bar z}w ]\}$$ 
$$ = {\mbox{e}^{-2pE} \over 1 + p^2E^2 -2pE-2pE^2/s + 2E/s},$$
for $p < E^{-1}$, whence 
\begin{equation} \label{51}
{\tilde \mu} (\pi, s, p) = s\{2pE + \log [1+p^2E^2 -2pE +
2E(1-pE)/s]\}.
\end{equation}

Trying to maximize $\mu (\pi, s, p)$ with respect to $p$ we obtain the
equation \begin{equation} \label{52}
(1 + E - pE)^s (1 - p) = E^s ,\end{equation}
which can be solved explicitly only for $s =0,1$. Thus, contrary to
the classical case \cite{gal}, the maximum in (\ref{47}) in general
can be found only numerically. For $s=0$ we have $p=0$ and
$$C = \frac{\partial}{\partial s}\mu (\pi, 0, 0) = (E+1)\log (E+1) -
E\log E.$$
For $s=1$ equation (\ref{52}) has the unique solution 
$p(1, E) = 1 + 1/E - g(E)/E < E^{-1}$, where 
$$g(E) = {1 + \sqrt{4E^2 + 1} \over 2}.$$
For future use we find the important quantities
$$
\mu (\pi, 1, p(1, E)) = 2(E + 1 - g(E)) + \log g(E);$$
\begin{equation} \label{53}
\frac{\partial}{\partial s}\mu (\pi, 1, p(1, E)) = E + 1 - g(E) + 
{g(E)^2\log g(E) -E^2 \log E \over g(E)^2 - E^2}. \end{equation}

The optimization of the expurgated bound can be
performed analytically. Taking partial derivative with respect to
$p$ we obtain the equation
$$p^2 - 2p\left({1 \over s} + {1 \over 2E}\right) + {1 \over sE} = 0,$$ 
the solution of which, satisfying $p < E^{-1}$, is
$$p(s, E) = s^{-1} + E^{-1} - E^{-1}g(E/s).$$ 
Substituting this in (\ref{48}), we obtain the following expression,
which is to be maximized with respect to $s\geq 1$:
$${\tilde \mu}(\pi, s, p(s, E)) - sR = 2(E + s - s g(E/s)) + s\log g (E/s)
- sR.$$ Taking derivative with respect to $s$, we obtain the equation
$$g(E/s) = \mbox{e}^R,$$ the solution of which is 
\begin{equation} \label{54} s = {E \over
\sqrt{\mbox{e}^{2R} - \mbox{e}^{R}}}. 
\end{equation}
If this is less than 1, which
is equivalent to $$R < \log g(E) = \frac{\partial}{\partial s}{\tilde
\mu} (\pi, 1, p(1,E)),$$ then   the maximum is achieved for the value of
$s$ given by (\ref{54}) and is equal to
$$2E (1 - \sqrt{1 - \mbox{e}^{-R}}) = E_{ex} (R) > E_r (R),$$
(which up to a factor coincides with the expurgated bound for
classical Gaussian channel). In the range 
$$ \frac{\partial}{\partial s}{\tilde
\mu} (\pi, 1, p(1,E)) \leq R \leq \frac{\partial}{\partial s}
\mu (\pi, 1, p(1,E)), $$where the optimizing $s$ is equal to 1, 
we have the linear bound$$E_{ex} (R) = E_r (R) = \mu (\pi, 1, p(1,E)) - R,$$
with the quantities $\frac{\partial}{\partial s}
\mu (\pi, 1, p(1,E)), \mu (\pi, 1, p(1,E))$ defined by
(\ref{53}). Finally, in the range $$\frac{\partial}{\partial s}
\mu (\pi, 1, p(1,E)) < R < C$$ we have $E_{ex} (R) < E_r (R)$ with $E_r
(R)$ given implicitly by (\ref{47}).

On the other hand, for the pure-state photon channel the analysis of
the error probability is trivial: since this is quasiclassical noiseless
channel, the error probability is zero for $R < C$. Thus, although the
two channels are asymptotically equivalent in the sense of capacity,
their finer asymptotic properties are apparently essentially different.
\vskip20pt
\centerline{\sc Acknowledgments}
\vskip10pt
The work was stimulated by discussions at the mini-workshop organized
by the Institute for Scientific Interchanges, Turin, February 1997.
The author is grateful to Prof. M. D'Ariano and Prof. M. Rasetti for the
opportunity to take part in this meeting. The author is also grateful to
Prof. H. P. Yuen for hospitality at Northwestern University, where
the work was accomplished. The work was partially supported by the RFBR
grant no. 96-01-01709.

\end{document}